\documentclass[a4paper]{jpconf}
\usepackage{graphicx}
\usepackage{enumitem}
\usepackage{amsmath}
\usepackage{amssymb}
\usepackage{amsfonts}
\usepackage{dsfont}
\usepackage{cite}
\usepackage{xcolor}

\begin{document}
\title{Steps towards Lorentzian quantum gravity with causal sets}

\author{Astrid Eichhorn}

\address{CP3-Origins, University of Southern Denmark, Campusvej 55, DK-5230 Odense M, Denmark and\\
Institute for Theoretical Physics, University of Heidelberg, Philosophenweg 16, 69120 Heidelberg, Germany}

\ead{eichhorn@sdu.dk}

\begin{abstract}
Causal set quantum gravity is a Lorentzian approach to quantum gravity, based on the causal structure of spacetime. It models each spacetime configuration as a discrete causal network of spacetime points. As such, key questions of the approach include whether and how a reconstruction of a sufficiently coarse-grained spacetime geometry is possible from a causal set. As an example for the recovery of spatial geometry from discrete causal structure, the construction of a spatial distance function for causal sets is reviewed. 
Secondly, it is an open question whether the path sum over all causal sets gives rise to an expectation value for the causal set that corresponds to a cosmologically viable spacetime. 
To provide a tool to tackle the path sum over causal sets, the derivation of a flow equation for the effective action for causal sets in matrix-model language is reviewed. This could provide a way to coarse-grain discrete networks in a background-independent way. Finally, a short roadmap to test the asymptotic-safety conjecture in Lorentzian quantum gravity using causal sets is sketched.
\footnote{Based on a talk given at the Ninth International Workshop DICE2018 
 Castello Pasquini/Castiglioncello (Tuscany), September 17-21, 2018
Spacetime - Matter - Quantum Mechanics. 
}
\end{abstract}

\section{Lorentzian quantum gravity}
Causal set quantum gravity is an approach to the path integral over spacetimes. Unlike many other approaches to quantum gravity, it emphasizes the spatiotemporal nature of gravity, i.e., it is intrinsically Lorentzian, and cannot be defined in a Riemannian setting. Therefore its starting point is a manifold with a Lorentzian metric, $(\mathcal{M}, g)$. Yet, due to diffeomorphism invariance not all components of the metric are actually physical. Instead, the physical content of the metric consists of the causal structure (locally corresponding to the lightcone at a point), providing the partial order relation $\prec$ (``precedes") and a local scale (encoded in the conformal factor), i.e., the local volume element $\epsilon$. 
The order relation $\prec$ on physical spacetimes should satisfy 
\begin{enumerate}[label=(\alph*)]
\item transitivity $x\prec y, \, y\prec w\,\, \Rightarrow x \prec w$
\item acyclicity (no closed timelike curves) $x\prec y,\, y\prec x\,\, \Rightarrow x=y$,
\end{enumerate}
for spacetime points $x,y,w$.
As causal order does not relate points at spacelike proper distance, it has no Riemannian equivalent and generates a partial instead of a total order in the Lorentzian case. \\
A manifold endowed with a causal structure and a local volume element, $(\mathcal{M}, \prec, \epsilon)$ is physically equivalent to $(\mathcal{M}, g)$, but only keeps the physical information encoded in the metric $g$, \cite{hkm,mal}, making $(\mathcal{M}, \prec, \epsilon)$ an arguably more suitable starting point for quantum gravity than $(\mathcal{M}, g)$. It underlies the causal-set approach to quantum gravity \cite{Bombelli:1987aa}, see \cite{Sorkin:2003bx,Dowker:2005tz,Henson:2006kf,Henson:2010aq,Surya:2011yh,Dowker:aza} for reviews. Causal set quantum gravity is based on the additional assumption of spatiotemporal discreteness. Accordingly, a causal set is  a set of spacetime points, together with the order relation $\prec$, which in addition to the conditions (a) and (b) satisfies local finiteness
\begin{enumerate}[label=(\alph*),start=3]
\item $\forall x, y \in \mathcal{C}:\, {\rm card}\{z \in \mathbb{C}, s.t.\,\, x\prec z \prec y\} < \infty$.
\end{enumerate}
The latter is the discretization condition. One can impose it in two conceptually very distinct ways: One can view it as a central part of the definition of the physical theory, as is actually done in causal set quantum gravity. In this view, geometric structures at sub-discretization scales on the continuum manifold have no physical meaning, and are therefore not encoded in the corresponding causal set.
Alternatively, one could regard condition (c) as a \emph{regularization} that enables the study of the regularized gravitational path integral, e.g., using numerical simulations or matrix-model techniques. In the second setting, the continuum limit is actually the physically relevant limit, cf.~Sec.~\ref{sec:ASfromCS}. \\
\begin{figure}
\centering
\includegraphics[width=0.5\linewidth]{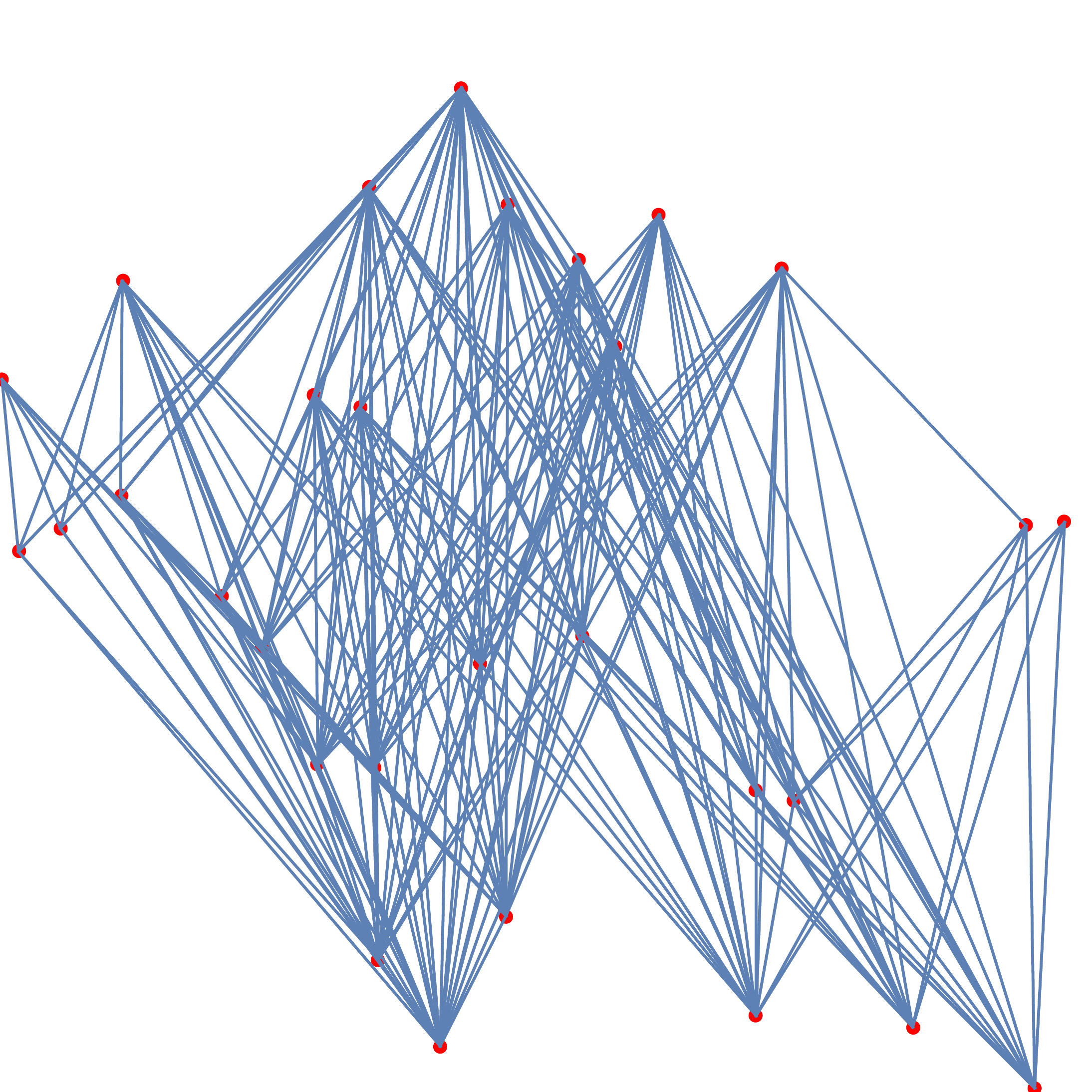}
\caption{\label{fig:sprinkling} A sprinkling into the interval $[0,1]\times[0,1]$ in $\mathds{M}^2$ with the red dots indicating the coordinates of the causal set elements in the embedding spacetime, and the blue lines indicating the existence of a causal relation. The embedding is not part of the fundamental definition of a causal set. }
\end{figure} 
The last missing ingredient to approximate $(\mathcal{M}, \prec, \epsilon)$ by a causal set is the discrete counterpart of the scale $\epsilon$:
A causal set approximated by a manifold encodes the information on the local scale in the number of elements: By associating a spacetime volume of Planckian size on average to each causal set element, the  number of elements provides the spacetime volume in Planckian units. This gives rise to the causal-set-``slogan" that ``order +number = geometry", coined by Rafael Sorkin, and underlies an argument that provided a prediction of the value of the cosmological constant \cite{lambdapaper}.
Specifically, a causal set is approximated by a manifold, if the causal set has a high probability to arise through a sprinkling process. A sprinkling is a random selection of spacetime points according to a Poisson distribution. The random nature of the sprinkling is crucial to avoid a violation of Lorentz invariance despite the discreteness.
An illustration of a causal set approximated by a region of 1+1-dimensional Minkowski spacetime is shown in Fig.~\ref{fig:sprinkling}.

\section{Key challenges of causal sets: Recovering geometry and making sense out of the path sum over all causal sets}

\subsection{Reconstructing geometry from causal structure}
The minimalistic structure of causal sets motivates the question whether a causal set can indeed encode all sufficiently coarse-grained geometric information of a continuum spacetime.
The Hawking-King-McCarthy/Malament-theorem indeed suggests that this could be the case. It applies to the continuum setting, and states that under certain causality conditions, the conformal geometry is completely determined by the causal space $(M, \prec)$. Therefore one could indeed expect that the causal relations together with the number-to-volume correspondence could be sufficient to recover a sufficiently coarse-grained geometry from a causal set. Explicit examples that information on continuum geometry can be reconstructed explicitly include the spacetime dimension \cite{Myrheim:1978ce,Meyer,Reid:2002sj,Glaser:2013pca}, timelike and spacelike distances \cite{Rideout:2008rk}, the spatial topology \cite{Major:2006hv,Major:2009cw,Sorkin:2018tbf}, the scalar d'Alembertian \cite{Benincasa:2010as,Aslanbeigi:2014zva,Belenchia:2015hca} and the scalar curvature in arbitrary dimensions \cite{Benincasa:2010ac,Dowker:2013vba}.

\begin{figure}
\centering
\includegraphics[width=0.5\linewidth,clip=true,trim=5cm 0cm 10cm 18cm]{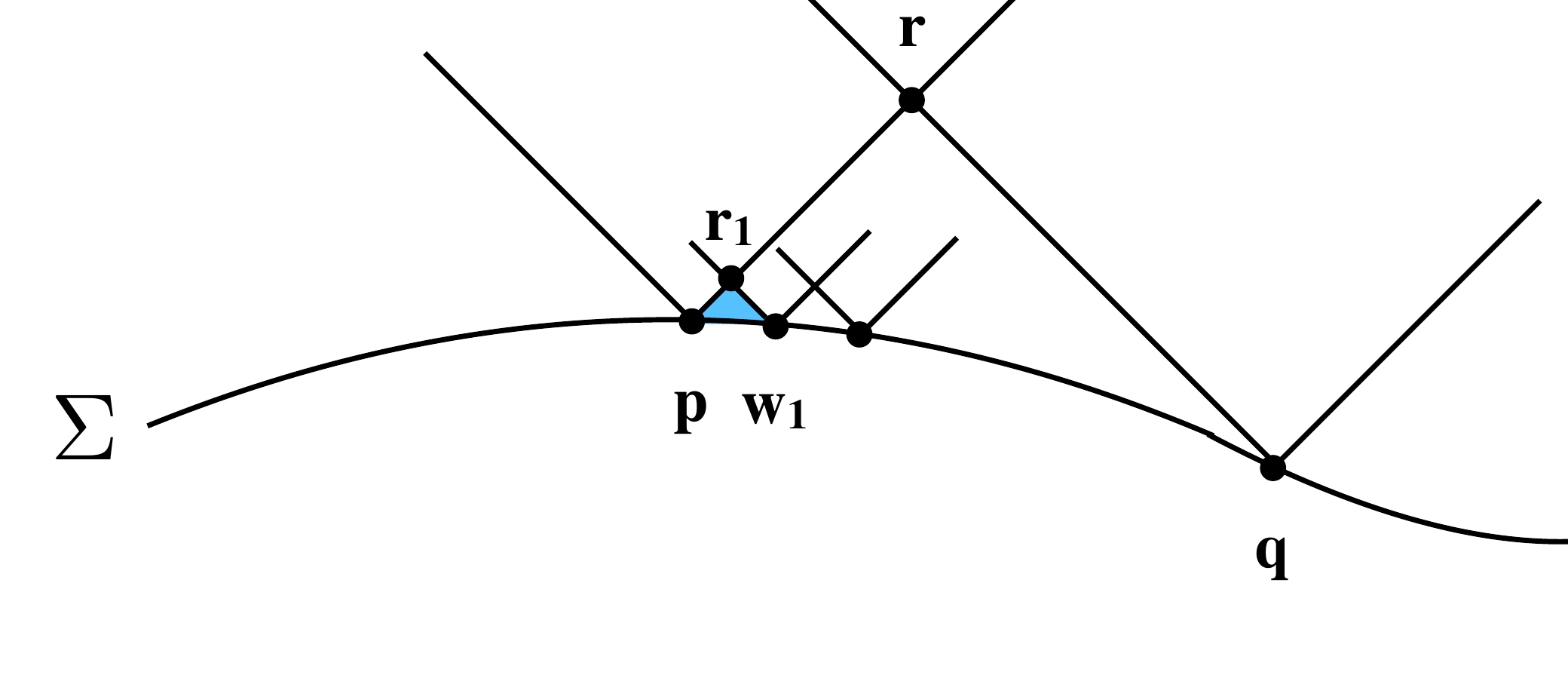}
\caption{\label{fig:ds}Illustration of an extrinsically curved hypersurface $\Sigma$ with the two points $\mathbf{p}, \mathbf{q}$. There distance is not well-approximated in terms of the suspended volume $V(r)$. Instead, a piecewise construction along a path $w_1...$ is done. For instance, the suspended volume $V(\mathbf{r}_1)$ (marked by the blue triangle) provides the correct distance between $\mathbf{p}, \mathbf{w}_1$.}
\end{figure}

 Due to the intrinsic spatiotemporal nature of causal sets one might expect the spatial geometry in a foliated spacetime to be most challenging to reconstruct. In \cite{Eichhorn:2018doy}, a proposal to extract the spatial distance between points in a causal set was put forward and tested numerically. In fact, in accordance with the  Hawking-King-McCarthy/Malament-theorem, the causal structure allows to recover the spatial geometry. The key idea of the construction for a continuum setting is the following: 

Given two points $p,q$ on a flat spacelike hypersurface $\Sigma$ in $\mathds{M}^d$, they have a common causal future, i.e., a set of spacetime points that is preceded by both $p$ and $q$. For every point $r$ in that set, its causal past $J^-(r)=\{s\vert s\prec r\}$ has a non-empty intersection with the future of $\Sigma$, $J^+(\Sigma)= \cup_{s\in\Sigma}J^+(s)$. This intersection is defined as $J(\Sigma,r)=J^+(\Sigma)\cap J^-(r)$ and has spacetime volume $V(r)={\rm vol}(J(\Sigma,r))$. The intersection $J(\Sigma,r)$ essentially constitutes an inverted cone, with the tip at the point $r$ and the base on the hypersurface $\Sigma$. This suspended spacetime volume provides a distance function on $\Sigma$. Specifically, the distance can be extracted from $V(\mathbf{r}_m)$, where $\mathbf{r}_m$ is the point that minimizes $V(r)$. 
For the case of $d$-dimensional Minkowski spacetime $\mathds{M}^d$ with a flat hypersurface $\Sigma$, the spatial distance between two points is accordingly given by
\begin{equation}
\widetilde{d}(p,q) = 2 \left(\frac{V(r)}{\zeta_d} \right)^{\frac{1}{d}},\label{eq:dV}
\end{equation}
where 
\begin{equation}
\zeta_d=\frac{\pi^{\frac{d-1}{2}}}{d\, \Gamma\left(\frac{d+1}{2}\right)}.
\end{equation}
This construction of the spatial distance makes no reference to the induced spatial metric on the hypersurface. Instead, it is extracted from the flat four-metric, by making use of the causal structure to the future of the hypersurface \footnote{An equivalent construction is possible by making use of the past of the spatial hypersurface $\Sigma$. In this case $\mathbf{r}_m$ constitutes the ``minimal" point in the common past of two points on $\Sigma$.}.\\
In the presence of intrinsic and/or extrinsic curvature, such a direct relation between the suspended volume and the distance no longer exists, as curvature corrections have to be taken into account. Yet, at distances sufficiently smaller than the smallest curvature scale $l_K$, where the metric becomes locally well-approximated by the Minkowski metric, the construction works. Therefore, a distance function between widely separated $p,q$ can still be defined using the relation Eq.~\eqref{eq:dV} by summing the distances in a piecewise fashion along a path between $p$ and $q$, and then minimizing over all possible paths, while limiting the maximum distance of each contributing piece. Specifically, we first define a predistance function $\widetilde{d}$ by using Eq.~\eqref{eq:dV}. In the presence of curvature, unlike in the case of a flat hypersurface embedded in $\mathds{M}^d$, this is not yet the distance.  For instance, in the presence of negative (positive) extrinsic curvature, $\widetilde{d}$ significantly underestimates (overestimates) the distance. In particular, this can lead to violations of  the triangle inequality.\\
Accordingly we construct the spatial distance in a piecewise fashion.
We define discretized paths in $\Sigma$, i.e., $W_k=(w_0,w_1,...,w_k)\subset \Sigma$, such that $w_0=p$, $w_k=q$. The path distance along such a path is defined as
\begin{equation}
d_{W_k}(p,q)= \sum_{i=0}^{k-1}\widetilde{d}(w_i,w_{i+1}).
\end{equation}
Herein, it is crucial to impose a mesoscale cutoff $\ell$, such that only such paths are allowed for which
\begin{equation}
\widetilde{d}(w_i,w_{i+1})<\ell.\label{eq:mesocutoff}
\end{equation}
The mesoscale cutoff has to be chosen sufficiently smaller than all curvature scales (extrinsic and intrinsic), i.e., $\ell \ll l_K$.
The distance function is the minimum of the path distance over all possible paths for which all steps satisfy Eq.~\eqref{eq:mesocutoff},
\begin{equation}
d(p,q) = \underset{W_k}{{\rm min}}\,\, d_{W_k}(p,q)
\end{equation}
 The construction is illustrated in Fig.~\ref{fig:ds}.

In summary, this provides a spatial distance purely from causal information, constituting one explicit example of the general insight of the Hawking-King-McCarthy/Malament-theorem.\\

This construction is easily adapted to the discrete causal-set case. The discrete analogue of a spatial hypersurface is an inextendible antichain $\mathcal{A}$, i.e., a set of unrelated elements, such that no further element can be added that is not related to an element already included in $\mathcal{A}$, \cite{Major:2005fy}. The suspended volume between a point in the future of the antichain and the antichain itself can be obtained by counting the number of elements in the corresponding causal interval. This allows to directly transfer the above definitions to the causal-set case: The predistance between two elements $p,q$ of the antichain is given by the minimum of the suspended volume, taken over all points in their common future $\mathcal{F}(p,q)$,
\begin{equation}
\widetilde{\bf{d}}(p,q) = \underset{\bf{e}\in \mathcal{F}(p,q)}{\rm min}2\left( \frac{\bf{V}(e)}{\zeta_d}\right)^{\frac{1}{d}}.
\end{equation}
Herein, $\bf{V}(e)$ is obtained by counting causal-set elements, and the dimension $d$ of the spacetime that the causal set embeds into must be known.

 In the antichain, there is a finite number of possible paths between two elements $p,q$. Specifically, one can form the set $\mathcal{W}(p,q)$ of all ordered subsets $\chi_k$ of $\mathcal{A}$ that contain $p$ and $q$, i.e., all paths between $p$ and $q$.
 The predistance can be calculated between all pairs of elements in any $\chi_k$. Summing these provides the predistance between $p$ and $q$ along a path $\chi_k$, directly mimicking the continuum construction:
\begin{equation}
d_{\chi_k}(p,q) = \sum_{i=0}^{k-1}\widetilde{\bf{d}}(e_i,e_{i+1}),
\end{equation}
where $\chi_k= \{p,e_1,...e_{k-1},q\}$.
 Subsequently, all paths are discarded where the predistance in at least one step is larger than the mesoscale cutoff as in Eq.~\eqref{eq:mesocutoff}, such that $\widetilde{\bf{d}}(e_i,e_{i+1})<\ell$.
The distance is defined by minimizing the path-distance over all remaining paths in the antichain $\mathcal{A}$, i.e.,
\begin{equation}
{\bf d}(p,q)= \underset{\chi \in \mathcal{W}(p,q)}{\rm min} d_{\chi_k}(p,q). 
\end{equation}
Here, it is crucial to impose the mesoscale cutoff on all pieces of the paths $\chi_k$, otherwise the minimization can result in violations of the triangle inequality.

In the limit of infinitely densely sprinkled causal-set elements, i.e., the discreteness scale going to zero, one obviously recovers the continuum construction and therefore a calculation of the spatial distance from the information on the causal structure. At finite discreteness scale, the interplay of  the mesoscale cutoff and the discreteness scale leads to an additional effect that is not present in the continuum.
\begin{figure}[!t]
\centering
\includegraphics[width=0.5\linewidth,clip=true, trim=5cm 0cm 0cm 6cm]{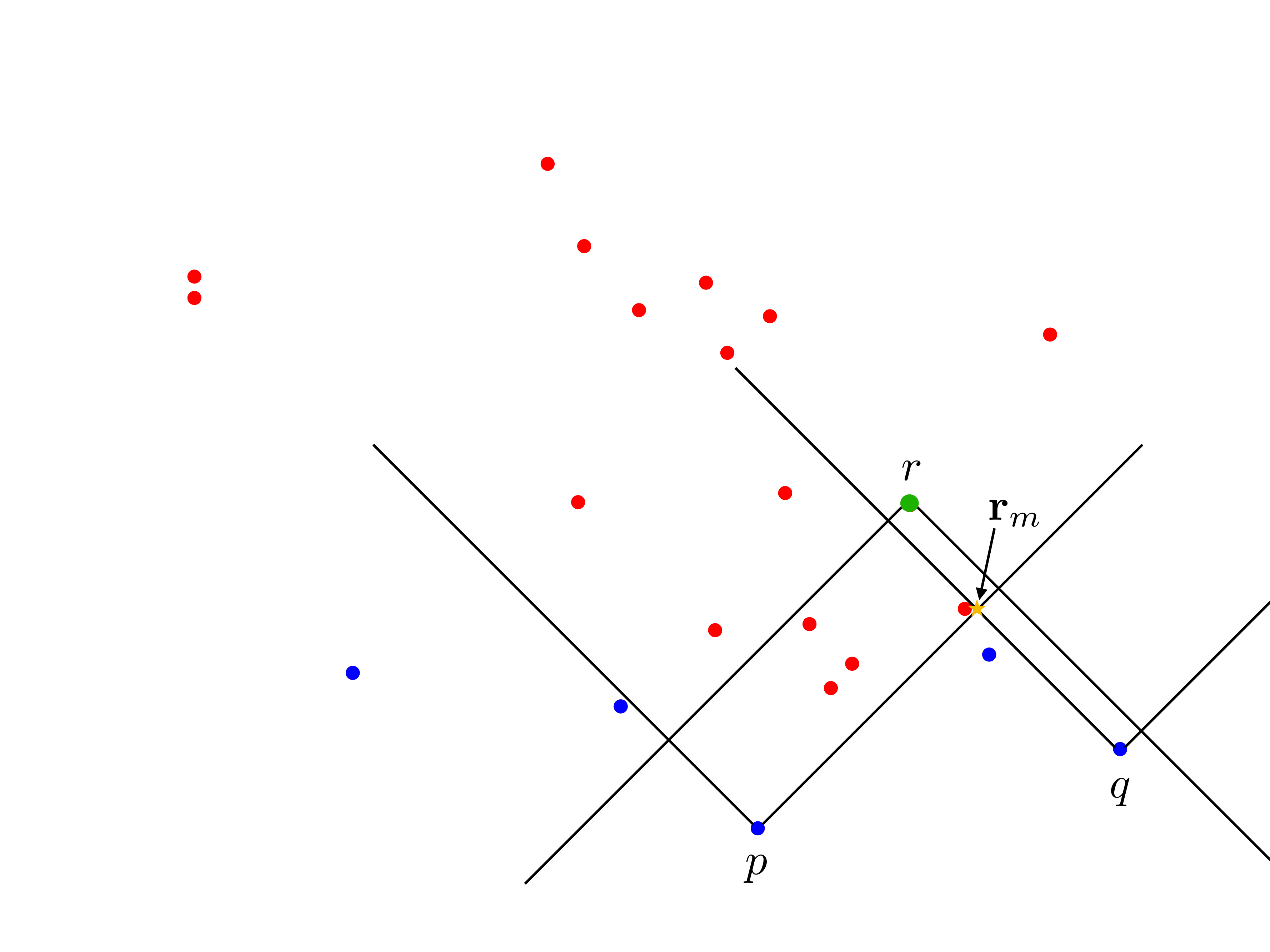}
\caption{\label{fig:DAS} The antichain at the bottom of the causal set shown in Fig.~\ref{fig:sprinkling} is marked in blue. In the continuum, the spacetime point $\mathbf{r}_m$, marked by a yellow star, would provide the distance between $p$ and $q$. There is no causal-set element at this point. Instead, the causal-set element marked in green at the spacetime point $r$ is the first in the common future of $p,q$. $V(r)$ is significantly larger than $V(\mathbf{r}_m)$, leading to an overestimation of the distance between $p$ and $q$.}
\end{figure}

\begin{figure}
\centering
\includegraphics[width=0.6\linewidth]{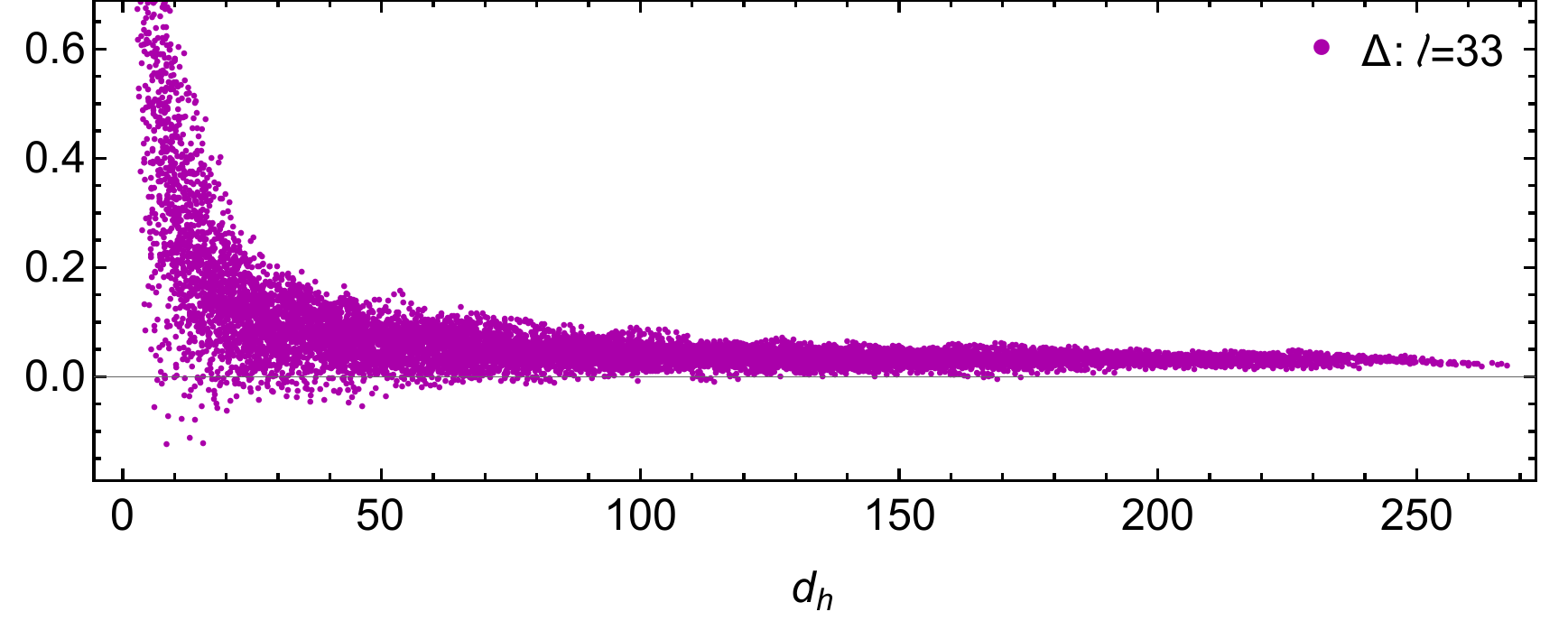}\\
\includegraphics[width=0.6\linewidth]{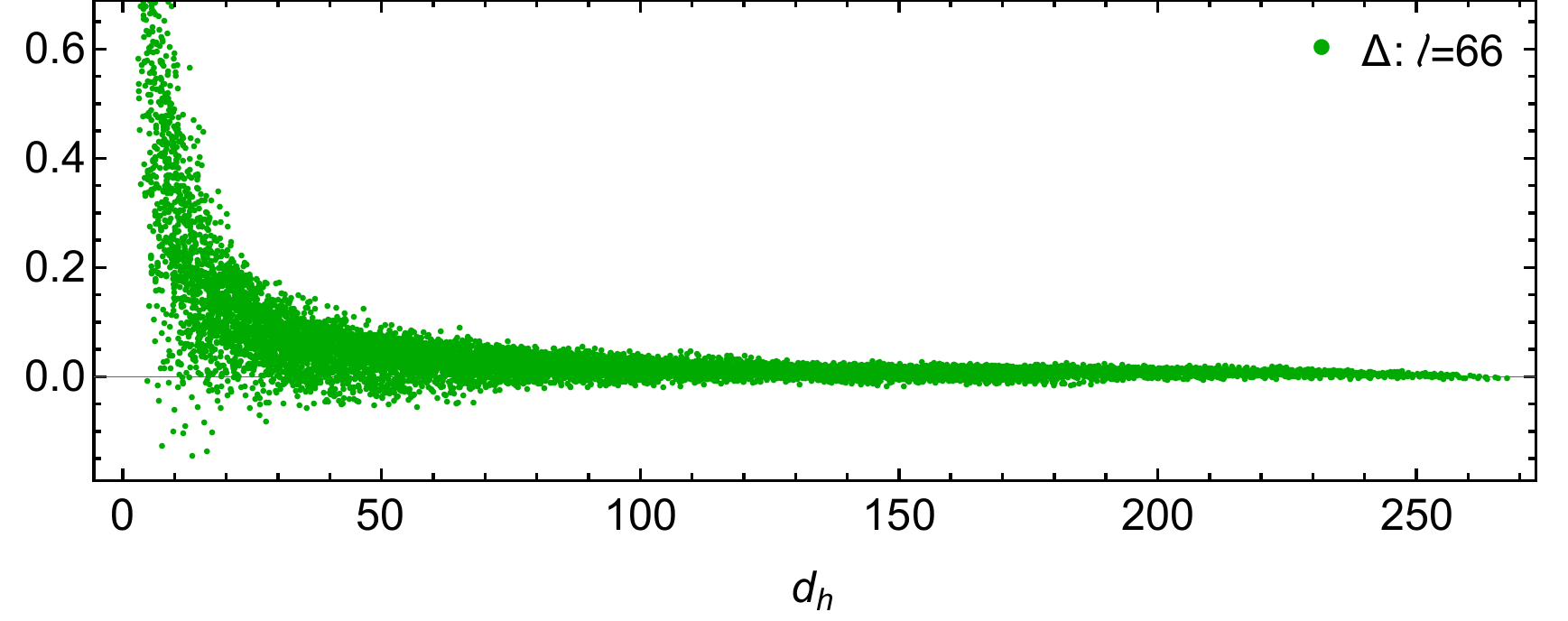}\\
\includegraphics[width=0.6\linewidth]{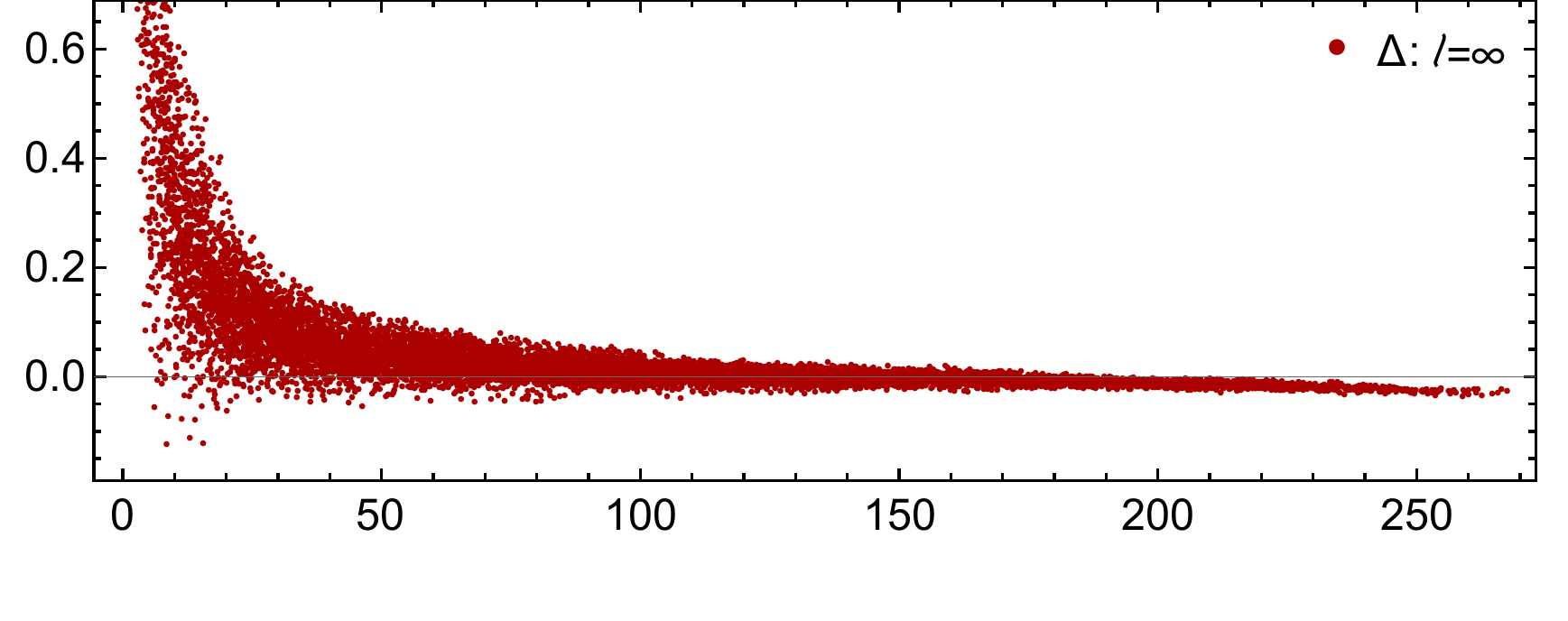}
\caption{\label{fig:simulations} To check how well the construction of the discrete distance ${\bf d}(p,q)$ compares to the continuum quantity $d_h(p,q)$ calculated in terms of the induced metric $h$ on $\Sigma$, we evaluate $\Delta = ({\bf d}(p,q)-d_h(p,q))/d_h(p,q)$ for sprinklings into $\mathds{M}^2$. In the above example, the hypersurface $\Sigma$ has constant extrinsic curvature characterized by the curvature scale $l_k=1000$, measured in units of the discreteness scale. For small distances $d_h \leq l_{DAS}$, the distance is overestimated, $\Delta>>0$. If the mesoscale cutoff is chosen too small, e.g., $\ell =33$ (upper panel), all distances are overestimated as a consequence of discrete asymptotic silence, $\Delta>0$ at all $d_h$. If the mesoscale cutoff is chosen too large, e.g., $\ell = \infty$ (lower panel), the distance is underestimated at larger distances due to the presence of the curvature scale. For this setting, the hierarchy of scales $l_{DAS}<<\ell<<l_{K}$, can be implemented, such that $\Delta \rightarrow 0$ for $d_h> l_{DAS}$ (middle panel). The data is taken from \cite{Eichhorn:2018doy}.}
\end{figure} 

This important difference between the discrete causal-set case and the continuum case is the presence of discrete asymptotic silence \cite{Eichhorn:2017djq,Carlip:2015mra}.
Given a causal set that is well-approximated by $\mathds{M}^d$, with an antichain that is well-approximated by a flat hypersurface and contains the points $p$ and $q$, the causal set does almost certainly not contain an element corresponding to the point $\mathbf{r}_m$, cf.~Fig.~\ref{fig:DAS}.  Instead, the first point $r$ in the common future of $p$ and $q$ lies ``further in the future", i.e., its suspended volume is larger than $V(\mathbf{r}_m)$. This effect can be understood as a form of decoupling of spatially nearby points: the causal interval relevant for the causal-set construction, $J(\Sigma,r)$ is larger than the causal interval that provides the distance in the continuum case, $J(\Sigma,\mathbf{r}_m)$. As this is similar to the effect of asymptotic silence \cite{Belinsky:1970ew} in classical gravity in the vicinity of spacetime singularities, the effect has been called discrete asymptotic silence. It is a direct consequence of the kinematically discrete nature of a causal set and vanishes in the limit of infinite sprinkling density.\\
The effects of discrete asymptotic silence set in at the length scale $l_{DAS}$ which lies about one or two orders of magnitude above the discreteness scale \cite{Eichhorn:2017djq}. Accordingly, the mesoscale cutoff $\ell$ should not be smaller than $l_{DAS}$, since distances $d \lesssim l_{DAS}$ are overestimated due to discrete asymptotic silence. If the piecewise distances along a path between $p$ and $q$ are overestimated due to this effect, then also ${\bf d}(p,q)$ will be overestimated. 

The construction of the spatial distance for causal sets accordingly requires a separation of scales according to
\begin{equation}
l_{DAS}<<\ell<<l_K.
\end{equation}
To determine how much of a separation is needed, numerical simulations are required. They indicate that $\ell\gtrsim 60\, \ell_{\rm disc}$, where $\ell_{\rm disc}$ is the discreteness scale, is sufficient. Numerical simulations in $\mathds{M}^{d}$ with $d=2,3$ with flat and extrinsically curved hypersurfaces in \cite{Eichhorn:2018doy} confirm that the causal information in a causal set is sufficient to reconstruct the spatial distance of points in the hypersurface, cf.~Fig.~\ref{fig:simulations}.

This provides another example that the intuition, based on the Hawking-King-McCarthy/Malament-theorem, that a causal set indeed encodes the  sufficiently coarse-grained geometry of a continuum spacetime, could be justified.

\subsection{The path integral for causal sets as a Renormalization Group equation} 
Recovering geometric information from appropriate causal sets and reconstructing a manifold is not sufficient to guarantee that the sum over all causal sets is well-defined and phenomenologically viable. This in particular requires that the expectation value taken over all causal sets provides a causal set approximated by a deSitter like spacetime, suitable for a description of our universe. A second challenge of the causal set program therefore consists in the evaluation of the path sum over all causal sets.
In particular, the set of all causal sets contains many causal sets which are not approximated by a Lorentzian manifold. Examples include the so-called Kleitman-Rothschild orders \cite{KR}, which consist of only three ``layers", with $N/4$ elements in the upper-most and lowest layer and $N/2$ elements in the middle. Each element in the middle layer has probability 1/2 to have share a link with each individual element in the upper and the lower layer. Clearly this structure cannot be embedded in an extended Lorentzian manifold. Entropically, i.e., based on a counting argument, these causal sets dominate the set of all causal sets starting already at relatively low numbers of elements \cite{Henson:2015fha}.
The path-sum over all causal sets is therefore not at all guaranteed to give rise to an effective Lorentzian spacetime manifold at a sufficiently coarse-grained scale. Moreover, the existence of non-manifoldlike causal sets also makes it more challenging to evaluate the path sum, e.g., using numerical simulations, as it very significantly enlarges the configuration space that must be sampled.

In \cite{Eichhorn:2017bwe} it was proposed to tackle the path integral over causal sets by writing it as a specific matrix model using the link matrix. Background-independent coarse-graining techniques for matrix models were introduced in \cite{Eichhorn:2013isa}, see \cite{Eichhorn:2018phj} for a review, and can be  applied to the causal-set case. More broadly speaking, this tool constitutes a way of coarse graining causal networks, irrespective of whether these are given a spacetime interpretation. As a first step, one rewrites causal sets as a matrix model: Each causal set comes with a matrix associated to it, the link matrix \cite{Johnston:2010su}. Links are the irreducible relations in a causal set. For two elements $e_i$, $e_j$, the corresponding entry in the link matrix is 1 if $e_i$ and $e_j$ share a link, i.e., $e_i\prec e_j$ and $\{e_k \in C\,|\,e_i\prec e_k\prec e_j\}=\o$. 
Otherwise the corresponding entry in the link matrix is zero.
The link matrix can be written down once the elements in the causal set are labelled. As the label cannot carry any physical meaning, different link matrices that arise by a relabelling of the same causal set are physically equivalent. The label independence is the discrete remainder of diffeomorphism invariance.  \\
In the path sum, one can thus sum over all link matrices (modulo relabellings) instead of all causal sets. To implement the path sum as a matrix model, also the action has to be written fully in terms of the elements of the link matrix. That this must be possible is clear, since the link matrix encodes the full structure of a causal set. To arrive at a background-independent dynamics, no other structure than the causal set itself can be used. Accordingly, nothing but the entries of the link matrix can occur in the dynamics. In particular, the number of $m$-paths, i.e., the number of paths with a total length of $m+1$ elements, between two elements $e_i$ and $e_j$ is encoded in $(L^m)_{ij}$. To respect background-independence and discrete diffeomorphism invariance, the full dynamics can only depend on $(L^m)_{ij}$. 
Specifically, one can also convince oneself explicitly that, e.g., the discrete analogue of the Einstein-Hilbert action can indeed be written purely as a function of the link matrix, e.g., the two-dimensional Benincasa-Dowker action \cite{Benincasa:2010ac} evaluated on a causal set $\mathcal{C}$ with $N'$ elements takes the form
\begin{equation}
S_{\rm BD}^{(2)}[\mathcal{C}] = N' -\sum_{i,j}\left(2L_{ij}- 4L_{ij}^2 +2\left(L_{ij}^3+\delta_{L_{ij}^2,2} \right)\right).
\end{equation}
The generating functional to be evaluated therefore takes the form
\begin{equation}
Z=\int \mathcal{D}L\, e^{iS[L]},
\end{equation}
where the sum goes over all link matrices of size $N'\times N'$, and the measure is understood to account for the discrete relabelling symmetry.
Introducing a source term $J_{ij}$ as well as a parameter $\alpha$ that allows for an analytical continuation, we define
\begin{equation}
Z[J]= \int \mathcal{D}L\, e^{-\alpha S[L]+ \sum_{i,j}J_{ij}L_{ij}}.\label{eq:defZ}
\end{equation}
The key idea of Renormalization Group techniques is to not perform this integral all at once but in steps. The decomposition into stepwise integrations should be done in such a way that the effective configurations that arise during the integration procedure capture physically relevant degrees of freedom. For instance, in local quantum field theories, one decomposes the integration into such steps that field configurations are averaged over in local patches, and long-wavelength fluctuations are integrated over in the very last steps. This provides coarse-grained effective configurations, which do indeed capture the relevant physics at large scales. In the absence of a background, which is the setting for several quantum gravity models, including causal sets, such a local coarse-graining cannot be defined. Instead, the number of degrees of freedom, i.e., the causal-set size, provides a scale according to which the integration can be done in a stepwise manner. This implements a coarse-graining procedure for causal sets that is rather intuitive: At the microscopic level, one starts with causal sets of size $N'$, and a corresponding microscopic dynamics. After a coarse-graining step, the same physics  is encoded in causal sets of size $N''<N'$ (i.e., \emph{coarser} causal sets), and an effective dynamics that includes the effect of the degrees of freedom that have been integrated out. This procedure can be implemented in the form of a formally exact equation that can be derived in analogy to a similar construction in the continuum \cite{Wetterich:1992yh,Ellwanger:1993mw,Morris:1993qb}: One generalizes the generating functional in Eq.~\eqref{eq:defZ} to a family of $N$-dependent generating functionals with $N<N'$ as follows:
\begin{equation}
Z_N[J] = \int \mathcal{D}L\, e^{-\alpha S[L] + \sum_{ij}J_{ij}L_{ij} + \frac{1}{2}\sum_{ijkl}L_{ij}R_N(i,j,k,l)L_{kl}},
\end{equation}
where 
\begin{eqnarray}
R_N(i,j,k,l) \,\,
\begin{cases}
>0\,\,\, {\rm for}\,\,\, (i+j+k+l)/N<1\\
 =0\,\,\, {\rm for}\,\,\, (i+j+k+l)/N>1.
 \end{cases}
\end{eqnarray}
This ensures that only entries of the link matrix with large index value are summed over, as the path sum over these entries is unchanged. In contrast, the finite value of $R_N$ for the low-index entries strongly suppresses their contribution in the path sum. Note that according to this definition, $N$ is an infrared cutoff scale in the generating functional, whereas $N'$ is a UV cutoff.
In a next step, it is convenient to express the dynamics in terms of the scale-dependent effective action, which is defined via a (modified) Legendre transform, 
\begin{equation}
\Gamma_N[\bar{L}] = \sum_{ij}J_{ij}L_{ij}- {\rm ln}Z_N[J] - \frac{1}{2}\sum_{ijkl}\bar{L}_{ij}R_N(i,j,k,l)\bar{L}_{kl},\label{eq:GammaN}
\end{equation}
where $\bar{L}_{ij}= \langle L_{ij}\rangle$ is the expectation value of the link matrix, i.e., the ``classical" link matrix.
The advantage of the transition to the scale-dependent effective action is that a differential equation can be derived that encodes the change of $\Gamma_N$ purely in terms of the second derivative of $\Gamma_N$, and no longer makes any reference to the microscopic action $S$.
The derivation makes use of 
\begin{equation}
\sum_{k,l} \left(\langle L_{ij}L_{kl}\rangle_N - \bar{L}_{ij}\bar{L}_{kl} \right) \left(\frac{d^2\Gamma_N}{dL_{kl}dL_{mn}}+R_N(k,l,m,n) \right) = \delta_{im}\delta_{jn},\label{eq:Gamma2}
\end{equation}
which says that the (regularized) propagator of the theory is related to the inverse of the second derivative of the scale-dependent effective action. Taking a derivative of the effective action in Eq.~\eqref{eq:GammaN} and using Eq.~\eqref{eq:Gamma2} results in
\begin{equation}
\partial_N\Gamma_N = \frac{1}{2}\sum_{ijkl} \partial_N R_N(i,j,k,l)\left(\frac{d^2\Gamma_N}{dL_{kl}dL_{mn}}+R_N(k,l,m,n) \right)^{-1}.\label{eq:floweq}
\end{equation}
This equation encodes how the effective dynamics changes as one integrates out entries in the link matrix, such that the physics that is encoded in the family of effective actions $\Gamma_N$ stays the same, even though the degrees of freedom become coarser in each step. This is just the usual idea underlying the Renormalization Group, adapted to a setting which lacks a local notion of scale, and instead comes with a background-independent notion of scale, namely $N$, the number of degrees of freedom.\\
Note that Eq.~\eqref{eq:floweq} is a formal equation which has not been tested in practical calculations yet. In the following we briefly discuss to what end studies based on Eq.~\eqref{eq:floweq} could be useful.\\
 Plugging in a form of the microscopic dynamics provides the initial condition that is necessary to integrate Eq.~\eqref{eq:floweq} and arrive at $\Gamma_{N\rightarrow 0}$. This is the full effective action that encodes the information on all correlation functions, and most importantly provides the quantum equation of motion
\begin{equation}
\frac{d\Gamma_{N\rightarrow 0}}{d \bar{L}_{ij}}=0.
\end{equation}
The solution provides the physical expectation value of $L_{ij}$. For causal set quantum gravity to be a candidate for a description of quantum spacetime in our universe,  the expectation value should be a sprinkling into a Friedmann-Robertson-Walker spacetime with a tiny cosmological constant in units of the Planck mass. Eq.~\eqref{eq:floweq} could be useful to tackle this question, because it could allow to calculate (an approximation to) $\Gamma_{N\rightarrow 0}$ in practise. While the full calculation of the effective action amounts to solving the full path sum over all causal sets, Eq.~\eqref{eq:floweq} provides a way of tackling this by studying a differential equation (instead of a sum/integral). Specifically, starting with a form of the dynamics at some initial scale, in addition to a scale-dependence of the couplings in this dynamics, new interaction terms are expected to be generated. Approximations to $\Gamma_{N\rightarrow 0}$ are obtained by taking these terms into account in successive enlargements of the truncation, starting by keeping only those interaction terms that are already present in the microscopic dynamics. Whether such truncations converge relatively quickly in the case of causal sets is presumably linked to the question whether the coarse-graining procedure reviewed above is based on physically relevant degrees of freedom. 

\section{Discreteness, the continuum limit and asymptotic safety in causal sets}

\subsection{Discreteness in quantum gravity}
As already highlighted, the discretization condition (c) can be understood either as part of the physical definition of the theory or as a regularization condition. The latter appears to contradict the commonly made assumption that quantum spacetime should be discrete. In particular, there are arguments, e.g., based on the finiteness\footnote{Note that the entanglement entropy in QFT across the horizon is actually finite once the scale-dependence of the Newton coupling is accounted for, see, e.g., \cite{Susskind:1994sm,Jacobson:1994iw,Demers:1995dq,Larsen:1995ax,Pagani:2018mke}} of black-hole entropy \cite{Sorkin:2005qx} as well as on the expected  production of black holes in the measurement of Planckian distances, that quantum gravity should come with a discreteness scale which should have the simultaneous interpretation of a minimum length/mininum volume, see \cite{Hossenfelder:2012jw} for a review. Sometimes the presence of Landau poles in the Standard Model of particle physics is advanced as an additional argument for the existence of a fundamental minimal length scale, although this is of course only one possible manifestation of the new physics that is required to render the Standard Model ultraviolet complete.
Based on this collection of arguments one might be led to expect that configurations of quantum spacetime must necessarily be discrete, and continuum approaches to quantum spacetime are automatically ruled out.  This would imply that the continuum limit in causal sets cannot have a physical meaning, and is at best of purely mathematical interest.

Yet, discreteness in quantum gravity could arise in different ways. In particular, models defined in the continuum (limit) could nevertheless exhibit physical discreteness. In other words, there is no automatic connection between kinematical and physical discreteness. This is also highlighted by models which are discrete at the kinematical level but nevertheless exhibit continuous dynamics. \\
An example for the first point is Loop Quantum Gravity, where the physical limit is the continuum limit. Nevertheless, the area and spatial volume operators exhibit discrete spectra, at least in the kinematical Hilbert space \cite{Rovelli:1994ge,Ashtekar:1996eg,Ashtekar:1997fb}.\\
 Moreover, one might at a first glance not expect that a model that is defined in the continuum can exhibit a minimal length scale. Asymptotically safe quantum gravity could constitute a counterexample: It exhibits a transition scale $l_{\rm tr}$, and is scale invariant at all distance scales $l/l_{\rm tr}<1$. This implies that all dimensionful quantities, such as the Planck length, scale canonically in that regime.  Accordingly, the Planck length acts as a particular form of minimum length scale \cite{Percacci:2010af}, as the ratio of the ``resolution" scale $l$ to the Planck length remains constant as  $l$ is lowered in the regime $l<l_{\rm tr}$. \\
Examples for continuous dynamics in discrete settings are given by perfect actions. These mimick continuum dynamics on a lattice even at finite lattice spacing. While these have been pioneered for non-Abelian gauge theories, \cite{Hasenfratz:1993sp}, their application in a quantum-gravity context has been discussed, e.g., in \cite{Bahr:2009qc}. In both cases, even though the model is kinematically discrete (i.e., each field configuration is discrete), the physics appears continuous.

Lastly, discrete configurations, i.e., kinematically discrete models could give rise to a continuum dynamics by being summed over in the path integral. For instance, in spin foams, this idea goes under the slogan that ``summing = refining" \cite{Rovelli:2010qx}.

These examples highlight that requiring a theory to show geometric discreteness (or conversely a geometric continuum) at the level of its physics content does \emph{not} imply that there must be discreteness (a continuum) at the kinematical level. Accordingly, even if we aim at developing a theory of quantum gravity that shows geometric discreteness, we can still choose a route via the continuum limit and view kinematical discreteness as a regularization. 

In particular, simple discrete models, where discreteness is interpreted as a cutoff to be imposed on configurations \emph{and} dynamics could suffer from a lack of predictivity near the cutoff scale. This is similar to a generic effective field theory, the dynamics of which are only constrained by symmetry, and accordingly contain infinitely many terms. In the infrared, all but finitely many of these terms are suppressed and have a vanishing impact on measurable quantities. Yet, at high energy scales, all infinitely many terms contribute. Therefore, infinitely many parameters, in the form of the couplings of the infinitely many interaction monomials, enter the dynamics. Then, one needs to impose a principle that provides relations between these couplings, such that only a finite number of free parameters remain. Demanding the existence of a universal\footnote{Universality in this context means that the continuum physics does not depend on the details of the regularization/discretization.} continuum limit is one possible way of achieving this. The reason lies in the fact that a universal continuum limit can be taken at a fixed point of the Renormalization Group flow: At such a fixed point, scale invariance is realized, allowing to ``zoom in" to arbitrarily small distances. Following RG trajectories away from the fixed point (towards finite cutoff scales) is only possible along finitely many directions in coupling space. Accordingly a model which allows one to take the continuum limit is (typically) specified by only a finite number of free parameters. 
Conversely, if discreteness is from the outset imposed on configurations and dynamics (as opposed to being emergent in a mathematically continuous setting), then any choice in this infinite-dimensional space of models is a priori viable for the dynamics at the cutoff scale. Just as in the effective field theory setting, predictivity is (expected to be) restored at scales far below the cutoff, where most couplings are suppressed in a canonical power counting \footnote{Whether this expectation from local QFTs also holds in, e.g., the non-local causal-set setting is an open question.}. Yet, close to the cutoff scale -- which in quantum gravity one would usually associate with the Planck scale -- predictivity is lost, unless a new principle is found that can be imposed to provide infinitely many relations between the various couplings.\\
For causal sets, discreteness is imposed on the configurations. Yet, the question whether the physics arising from causal sets will exhibit a discreteness scale is not yet clarified. For instance, just as in spin foams, the summation over all discrete configurations might potentially even give rise to continuum physics. Conversely, it is not excluded that it is actually necessary to demand the existence of a continuum limit to render the path integral for causal sets predictive and fix the freedom in the choice of microscopic dynamics. This necessitates the search for a Renormalization Group fixed point in the setting where all causal sets, also all non-manifoldlike ones, are taken into account. As argued above, this does not a priori preclude the emergence of physical discreteness, e.g., in the form of discreteness at the level of geometric observables.

\subsection{Sketch of a roadmap towards Lorentzian asymptotic safety from causal sets}\label{sec:ASfromCS}
In what follows, we view causal sets not as ``fundamental" configurations of spacetime, but rather as regularizations of the underlying continuum geometry. This differs from the discussion in the previous subsection, as here we focus only on the path sum over sprinklings and exclude, e.g., KR-orders, as these are not discretizations of Lorentzian manifolds and accordingly not part of the relevant configuration space.

This setup provides a novel angle from which to explore the asymptotic-safety conjecture for quantum gravity \cite{Weinberg:1980gg,Reuter:1996cp}. The asymptotic-safety program proposes that the path integral for local continuum quantum gravity could make sense at and beyond the Planck scale, where the usual effective-field theory treatment breaks down. The extension beyond the Planck scale becomes possible because of the interacting Reuter fixed point, which underlies a scale-invariant regime that is realized asymptotically. Intuitively speaking, this allows us to ``zoom in" into the dynamics up to arbitrarily small scales without running into any inconsistencies.
Compelling indications for the Reuter fixed point exist in a Riemannian setting, \cite{Reuter:2001ag,Litim:2003vp,Codello:2008vh,Benedetti:2009rx,Falls:2013bv,Dona:2013qba,Becker:2014qya,Gies:2016con,Denz:2016qks,Christiansen:2017bsy}, see \cite{Niedermaier:2006wt,Reuter:2012id,Percacci:2017fkn,Eichhorn:2017egq,Eichhorn:2018yfc} for reviews and \cite{Wetterich:2019qzx} for a recent account of quantum scale invariance. As no straightforward Wick rotation is available in quantum gravity, this does not imply the existence of an analogous fixed point in the Lorentzian case. Moreover, important questions such as whether quantum gravity breaks global symmetries due to the presence of virtual black-hole configurations \cite{Kallosh:1995hi}, as well as whether there is a UV-IR-interplay through black-hole formation, cannot be conclusively tackled in the Riemannian case, as only the spacetime exterior to the black-hole horizon can be continued analytically, see, e.g., \cite{Baldazzi:2018mtl}. Accordingly, exploring asymptotic safety in a Lorentzian setting is a key outstanding challenge. In continuum functional RG studies, this is technically challenging; for a first study see \cite{Manrique:2011jc}. In \cite{Eichhorn:2017bwe}, it was instead proposed that causal sets could serve as a framework to search for Lorentzian asymptotic safety. 
Specifically, sprinklings provide discretizations of all configurations of spacetime that enter the path integral. Accordingly, a regularized form of the Lorentzian path integral over all metrics is provided by the path sum over all sprinklings $\mathcal{C}_{sp}$, i.e., 
\begin{equation}
 \int \mathcal{D}\mathcal{C}_{sp}\, e^{i\,S^{(d)}[\mathcal{C}_{sp}]}\rightarrow \int \mathcal{D}g_{\mu\nu}e^{i\, S[g_{\mu\nu}]},
\end{equation}
where $S^{(d)}[\mathcal{C}_{sp}]$ is a $d$-dimensional microscopic dynamics selected by the asymptotic-safety conjecture.
 If the Lorentzian path integral over all metrics is well-defined, then the continuum limit can be taken if $S^{(d)}[\mathcal{C}_{sp}]$ is used to determine the phase of configurations. This is expected to be equivalent to asymptotic safety, since the perturbative non-renormalizability of gravity prohibits a well-defined continuum limit at the free fixed point. Therefore, searching for the existence of a continuum limit in the path sum over all sprinklings provides insight into whether asymptotic safety is realized in Lorentzian gravity.
 
To search for such a universal continuum limit, one has (at least) two options. The first is to use the Renormalization Group equation discussed above. In analogy to matrix and tensor models, one might expect that a universal continuum limit is visible as a fixed point in the large-$N$ limit. A fixed point at an arbitrary point in the space of couplings can be searched for with the flow equation Eq.~\eqref{eq:floweq}, as it does not make any reference to a microscopic action any more. Instead, the flow equation encodes a local vector field in the space of couplings, as becomes clear by expanding $\Gamma_N$ in terms of the couplings and the corresponding interaction monomials. A zero of this vector field corresponds to a fixed point.
This fixed point ultimately determines $S^{(d)}[\mathcal{C}_{sp}]$ and can a priori lie anywhere in the space of couplings. Accordingly, for a search for Lorentzian asymptotic safety with causal sets it is most likely not sufficient to map out the phase diagram in the space spanned by the Newton coupling and the cosmological constant, i.e., the two couplings of the classical Einstein-Hilbert action. Instead, one is presumably required to include further couplings, corresponding to higher-order terms in the gravitational dynamics. For instance, the construction in \cite{Benincasa:2010ac} can presumably be generalized to provide the causal-set analogue of $\int d^dx\, \sqrt{-g}\left(R^2+ \Box R\right)$ as well as higher orders in the curvature. As a first step towards asymptotic safety, one could simply work in a theory space spanned by all possible functions of the elements of the link matrix. Their geometric interpretation need not be known in order to determine whether Lorentzian gravity is asymptotically safe. 

An important advantage of causal sets in contrast to a discretization in the form of a regular lattice is that a random sprinkling does not break Lorentz-invariance as it does not select a preferred frame. This presumably has important implications for the continuum limit, as typically it requires the tuning of additional parameters to recover symmetries broken by a lattice discretization. In a Renormalization-Group language, the breaking of symmetries leads to an enlarged theory space in which fixed points acquire additional relevant directions. 
 
 As an alternative tool to search for a universal scaling regime, one can evaluate the path sum over all sprinklings by Monte Carlo simulations. Just as the use of the flow equation, this requires to introduce a parameter in the generating functional that allows for an analytical continuation.
The search for a universal continuum limit proceeds by mapping out the phase diagram in terms of the microscopic couplings. The existence of a universal continuum limit implies the existence of a second-order phase transition in the space of couplings. In the vicinity of a second-order phase transition, a universal scaling regime sets in, linking the physics in the vicinity of the phase transition to fixed-point behavior. \\
Although the corresponding studies are based on a different configuration space, it can be viewed as an encouraging result that \cite{Surya:2011du} finds a second-order phase transition. This would enable a continuum limit in the restricted class of two-orders, which are a subset of all causal sets. Moreover, the addition of a simple toy model of matter leads to additional second-order phase transitions \cite{Glaser:2018jss}. Extensions of these results to larger classes of causal sets might be feasible making use of the algorithms developed in \cite{Cunningham:2017pje}.
First steps along a similar direction, exploring asymptotic safety numerically using discrete random graphs have been made in \cite{Kelly:2018diy}.
 
This example highlights that,  as in many other difficult questions in quantum gravity,  there could be significant untapped potential hiding in the unexplored connections between different approaches to quantum gravity.

\subsection{Acknowledgements}
I  thank R.~Sorkin, S.~Surya and F.~Versteegen for helpful discussions, S.~Surya and F.~Versteegen for collaboration on \cite{Eichhorn:2018doy}, and the organizers of the Ninth International Workshop DICE2018 
 Castello Pasquini/Castiglioncello (Tuscany), Spacetime - Matter - Quantum Mechanics for the invitation.
I acknowledge support by the DFG through an Emmy-Noether fellowship under grant no.~Ei/1037-1.

\end{document}